\documentclass[12pt,letterpaper]{article}

\usepackage{graphicx}        
\usepackage{color}           
\usepackage{breakurl}        
\usepackage{natbib}
\usepackage{subcaption}
\usepackage{caption}

\usepackage{authblk}

\newcommand{\degree}{$^{\circ}$}

\newcommand{\mmss}{\mathrm{mm}/\mathrm{s}^{2}}

\chardef\us=`\_

\begin{document}


\title{The High Inclination Solar Mission}

\author[1]{Kobayashi, Ken}
\author[1]{Johnson, Les}
\author[1]{Thomas, Herbert D.}
\author[2]{McIntosh, Scott}
\author[1]{McKenzie, David}
\author[3]{Newmark, Jeffrey}
\author[1]{Heaton, Andy}
\author[1]{Carr, John}
\author[1]{Baysinger, Mike}
\author[1]{Bean, Quincy}
\author[1]{Fabisinski, Leo}
\author[1]{Capizzo, Pete}
\author[1]{Clements, Keith}
\author[1]{Sutherlin, Steve}
\author[1]{Garcia, Jay}
\author[4]{Medina, Kamron}
\author[4]{Turse, Dana}

\affil[1]{NASA Marshall Space Flight Center}
\affil[2]{High Altitude Observatory}
\affil[3]{NASA Goddard Space Flight Center}
\affil[4]{Roccor, LLC}


\maketitle

\begin{abstract}

The High Inclination Solar Mission (HISM) is a concept for an out-of-the-ecliptic mission
for observing the Sun and the heliosphere.  The mission profile is largely based on the
Solar Polar Imager concept\citep{2008nssv.book....1L}; initially taking $\sim$2.6 yrs to spiral in to a 0.48 AU ecliptic
orbit, then increasing the orbital inclination at a rate of $\sim 10$ degrees per
year, ultimately
reaching a heliographic inclination of $>$75 degrees. The orbital profile is
achieved using solar sails derived from the technology currently being developed for the Solar Cruiser mission, currently under development.

HISM remote sensing instruments comprise an imaging spectropolarimeter
(Doppler imager / magnetograph) and a visible light coronagraph.
The in-situ instruments include a Faraday cup, an ion composition spectrometer,
and magnetometers. Plasma wave measurements are made with electrical antennas and
high speed magnetometers.

The $7,000\,\mathrm{m}^2$ sail used in the mission assessment is a direct extension of
the 4-quadrant $1,666\,\mathrm{m}^2$ Solar Cruiser design and employs 
the same type of high strength composite boom, 
deployment mechanism, and membrane technology.
The sail system modelled is spun (~1 rpm)  to assure required boom characteristics with margin.
The spacecraft bus features a fine-pointing 3-axis stabilized instrument platform that
allows full science observations as soon as the spacecraft reaches a solar distance of 0.48 AU. 

\end{abstract}

\section{Introduction}

The value of a high ecliptic inclination vantage point for solar and heliospheric
observations has long been recognized. To date, the only mission that provided
a vantage point from outside the ecliptic plane was
the Ulysses mission \citep{1992A&AS...92..207W}.
However, Ulysses' payload was limited to in-situ instruments, and 
because of the long period
orbits necessitated by the Jupiter gravity assist, it only made 3 orbits (3 passes above 
north and south poles of the Sun respectively).

Apart from Ulysses, the only spacecraft with non-negligible orbital inclination 
is the recently launched Solar Orbiter \citep{2005AdSpR..36.1360M}, which will reach a
heliographic inclination of 
24 degrees at the end of the 7-year mission, and 33 degrees with an additional
three years of extended mission. However, even the 33 degree heliographic inclination
is marginal for observing the high latitude structures through helioseismology, and
inadequate for observing the solar wind from polar regions.

Several concepts have been put forward for out-of-the-ecliptic missions,
some using solar sails.
The Solar Polar Imager\citep{2008nssv.book....1L} (SPI) or the POLARIS\citep{POLARIS}
concept proposed a solar sail
mission with a 28,800 m$^2$ sail and 325 kg payload, achieving a final orbit of
0.48 AU and 75 degree inclination. SPI is specifically endorsed by
the 2013 Decadal Strategy for Solar and Space Physics (Heliophysics) \cite{NAP13060}.
Other proposals include The POLAR investigation of the Sun {\textemdash}POLARIS \citep{POLARIS} and Solar Polar Diamond Explorer (SPDEx) \citep{2018arXiv180504172V}.
But these concepts require significant technological development. 

Since the original SPI concept was originally 
formulated, there has been significant development
and testing in solar sail technology. Japan Aerospace Exploration Agency
(JAXA) successfully
deployed the IKAROS solar sail technology \citep{Yamaguchi}, and The Planetary Society
also demonstated successful deployment of solar sails \citep{lightsail}. 
Marshall Space Flight Center (MSFC)
and the Jet Propulsion Laboratory (JPL) have developed the NEA Scout \citep{neascout},
currently scheduled for launch in 2021. Solar Cruiser is currently in phase-A study
at MSFC, and uses a 16,000 m$^2$ sail based on the 82 m$^2$ NEA Scout sail design.

The High Inclination Solar Mission (HISM) is a concept study conducted at the MSFC Advanced
Concepts Office (ACO), with the goal of extending the Solar Cruiser sail design to
build a solar and heliospheric mission.

\section{Science Rationale}

A high latitude imaging system - like HISM - offers an opportunity to routinely observe the Sun’s polar regions. From that unique vantage point a broad spectrum of scientific challenges can be addressed, from studies of the Sun’s internal circulation and convection, detailed measurements of the fast solar (polar) wind and its interface with that from the magnetically closed equatorial regions. Beyond that, a polar vantage point permits the first synchronic study in the development of longitudinal waves that form on the Sun's activity belts that shape space weather and give rise to the largest solar eruptions. 

\begin{figure}[!ht]
    \centering
    \includegraphics[width=1.0\textwidth]{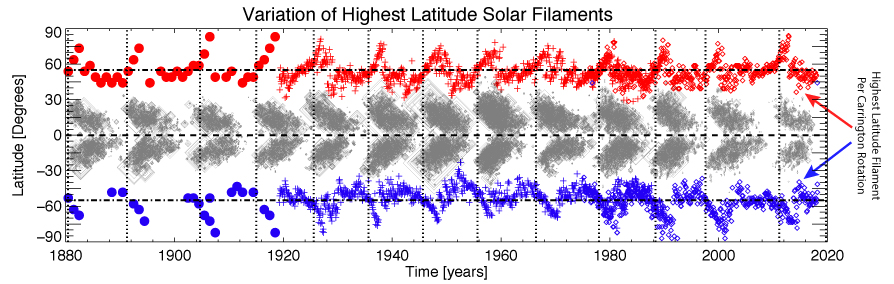}
    \caption{The 140 year record of solar filaments as observed in H${\alpha}$ observations from three sites: Arcetri Astrophysical Observatory (AO; 1880--1929), Meudon Observatory (MO; 1919\--1989), and the Kislovodsk Observatory (KO; 1980\--2018). Contrast the sunspot butterfly pattern in gray with the modulation of the highest latitude filaments present at each timestep (red:north; blue:south) and each observatory. In each figure we show the locations of the terminators as provided by \cite{2014ApJ...792...12M} as vertical dotted lines (see above). The dot\--dashed lines at $55^{\circ}$ in each hemisphere are added for reference.}
\label{fSMC1}
\end{figure}

The polar regions present tantalizing observational hints of the Sun’s cyclic behavior, and possibly that they play a critical role in establishing that critical behavior. Consider the composite of solar filament progression over the last 140 years constructed from a number of observatories \citep[][and Fig.~\ref{fSMC1}]{2019SoPh..294...88M}. Paying attention to the position and timing of the highest latitude filaments in the system we notice that \cite{2019SoPh..294...88M} spend much of their time around 55 degrees latitude, only progressing poleward on one occasion per solar cycle \-- at the start of the polar magnetic field reversal. This latitude is also the lowest of polar coronal holes \citep{2014ApJ...792...12M} and thus appears to present a robust canonical boundary between the polar and equatorial environments. {\it What is so special about 55 degrees?}

\begin{figure}[!ht]
    \centering
    \includegraphics[width=0.5\textwidth]{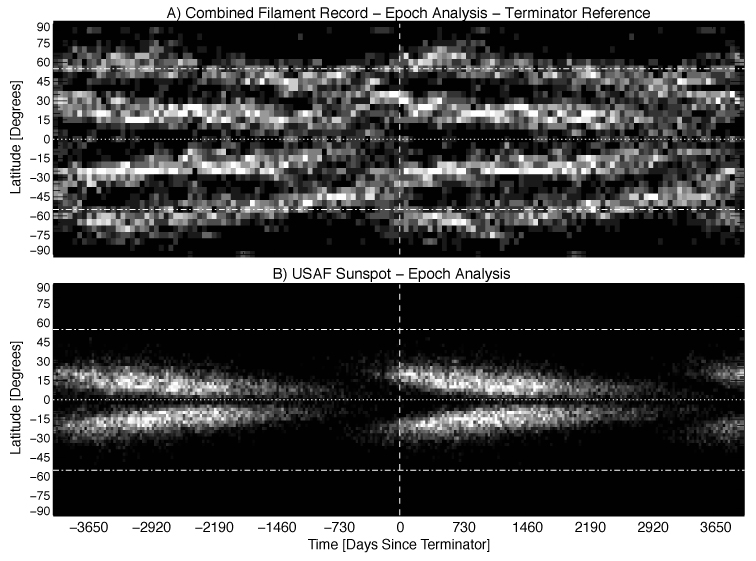}
    \caption{Comparing the SEA-derived mean patterns of filaments, and sunspots from 140 years of observations. The vertical dashed line indicates a time of zero. The horizontal dashed line indicates the equator while the horizontal dot\--dashed lines signify $55^{\circ}$ latitude in each hemisphere.}
\label{fSMC2}
\end{figure}

The vertical dashed lines in Fig.~\ref{fSMC1} represent the termination points of the magnetic activity bands that belong to the 22-year Hale magnetic cycle at the Sun’s equator \citep{2019SoPh..294...88M}. At those times, not only does the polar reversal process (REF) start, but we observe the birth of the sunspot cycle at mid-solar latitudes. Emphasizing the cyclic nature and importance of 55 degrees, we can construct a superposed epoch analysis (SEA) using the Hale cycle terminations as the key time. Figure~\ref{fSMC2} contrasts the SEA of 140 years of the filament density on the solar disk with the sunspot record over the same timeframe. That analysis shows that the sunspot progression is a subset of the Hale cycle manifestation that has its start at around 55 degrees around 10 years before. {\it Does this mean that the Sun’s dynamo process originates at 55 degrees and is more cyclic than the sunspot modulation would lead us to believe?}

The results shown above illustrate that imaging of the Sun’s high latitudes is possible when line-of-sight measurements become limited, they have been made possible by the cataloging of prominent features on the disk and above the limb. However, to probe and understand the physical processes taking place at 55 degrees that give rise to these features requires a study of the circulatory and convective flows and a high latitude observing platform. 

\begin{figure}[!ht]
\centering
\begin{subfigure}{.4\textwidth}
  \centering
  \includegraphics[width=\linewidth]{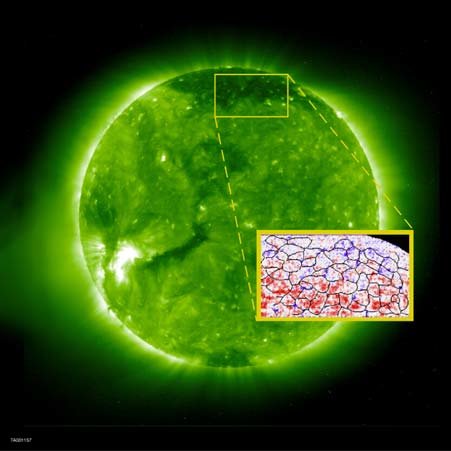}
  \label{fSMC4a}
\end{subfigure}
\begin{subfigure}{.4\textwidth}
  \centering
  \includegraphics[width=\linewidth]{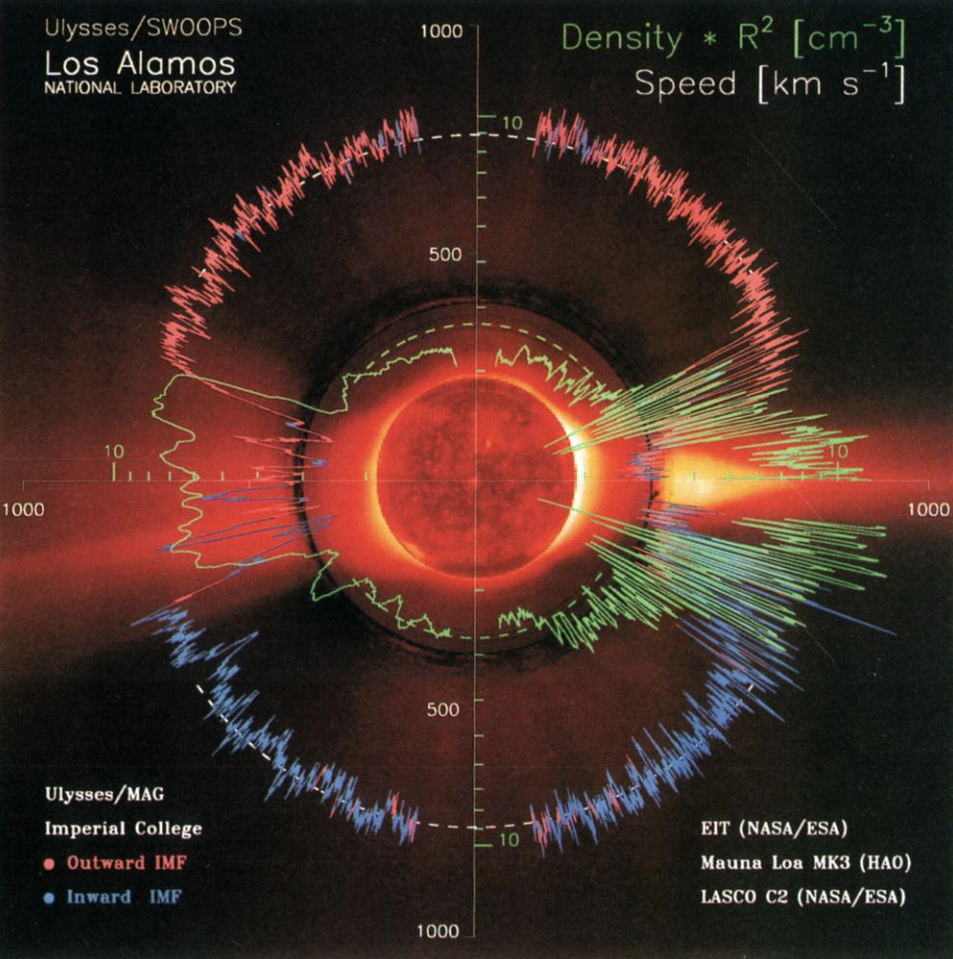}
  \label{fSMC4b}
\end{subfigure}
\caption{Left: The preferential plasma outflow visible in the ultraviolet spectral lines that form in the solar transition region at the supergranular network of polar coronal holes \-- the roots of the fast solar wind \citep[][]{1999Sci...283..810H}; Right: The measurements of Ulysses demonstrated the stark transition between the slow (magnetically closed) and fast (magnetically open) solar wind regimes \citep[][]{2000JGR...10510419M}.}
\label{fSMC4}
\end{figure}

Helioseismic measurements have probed the nature of the Sun’s interior and rotational characteristics for several decades. Ground-breaking inversions of line-of-sight Doppler measurements from the ecliptic have revealed an internal rotational structure that appears to be rigid below 70 percent of the Sun’s radius, but highly structured with depth in the outer 30\% \citep{1996Sci...272.1300T}. The left panel of Fig.~\ref{fSMC4} shows the state of the art in understanding the Sun’s internal rotation that, when presented in a slightly different format (right panel) shows that the inferred rotational profile with depth is uniform around 55\degree{} for a considerable depth (0.8 - 0.95 solar radii), hinting at the presence of a significant shear in the plasma flow there (possibly even an interface between the polar and equatorial regions). Analogous helioseismic investigations over the same epoch \citep[e.g.,][]{1980ApJ...239L..33H, 2000Sci...287.2456H} have revealed the presence of a banded flow pattern - of alternating fast and slow regions - that mirror the Hale magnetic cycle pattern deduced from surface features \citep[e.g.,][]{1987Natur.328..696S,1988Natur.333..748W,2014ApJ...792...12M}.
\begin{figure}[!ht]
\centering
\begin{subfigure}{.4\textwidth}
  \centering
  \includegraphics[width=\linewidth]{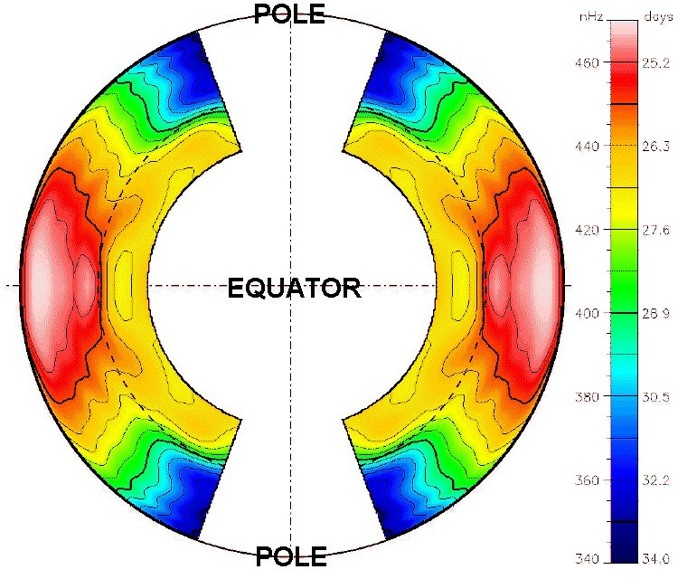}
  \label{fSMC3a}
\end{subfigure}
\begin{subfigure}{.5\textwidth}
  \centering
  \includegraphics[width=\linewidth]{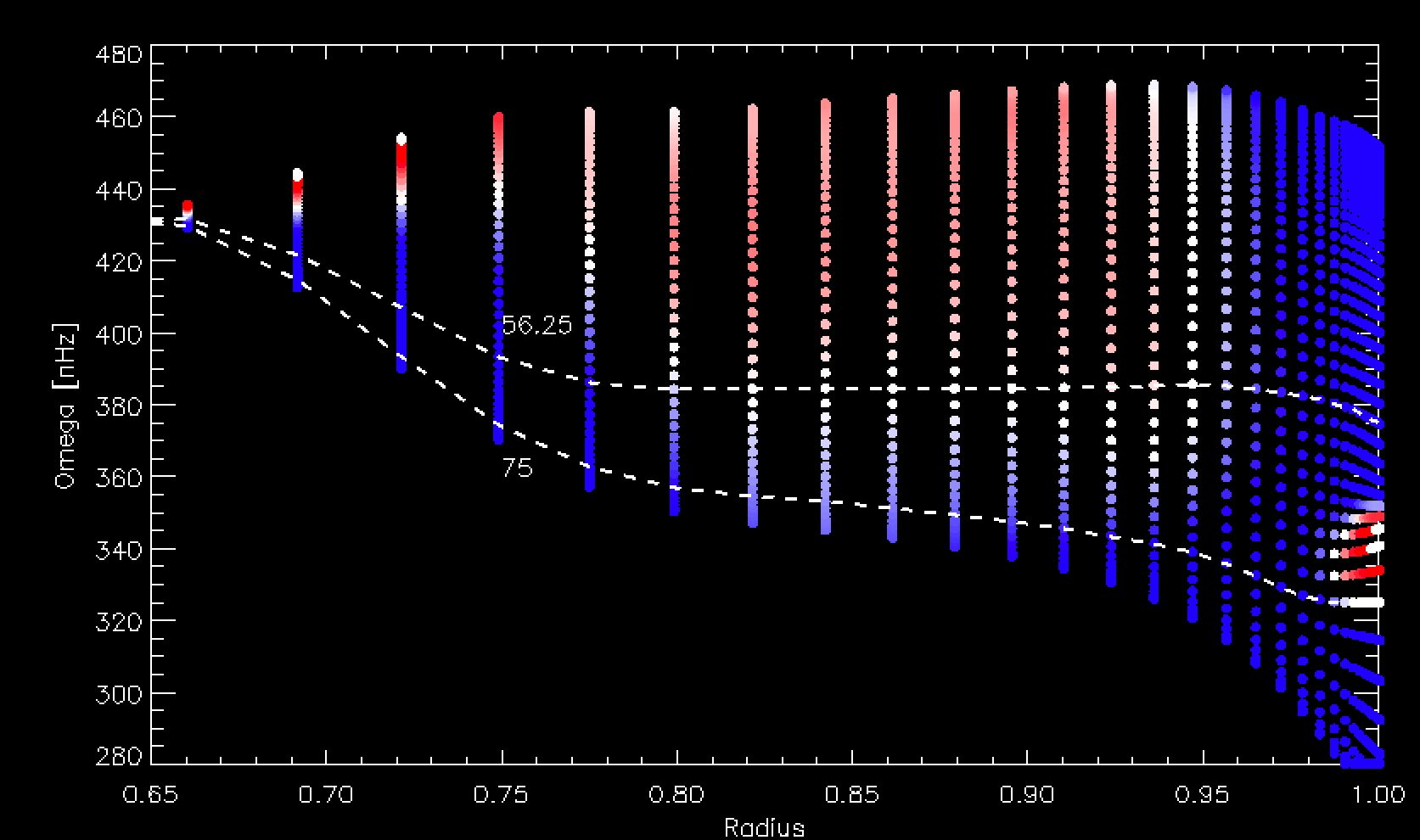}
  \label{fSMC3b}
\end{subfigure}
    \caption{Left: The rotation rate inside the Sun, determined by helioseismology using instruments aboard the showing the location of the tachocline where rigid rotation in the radiative zone gives way to differential rotation in the convective zone \cite{1996Sci...272.1300T}; Right: Higher spatial resolution version of rotation profile showing the average rotation profile (symmetrised) over the whole of the SDO/HMI record. Notice that near 55 degrees - the variation of the rotation rate with depth is close to zero.}
\label{fSMC3}
\end{figure}

From the particulate perspective of the solar wind the long periods of time where polar coronal holes are visible provide two unique scientific opportunities for a high-latitude platform: 1) the combined imaging and in-situ study of fast solar wind direct from its origin as inferred from Ulysses and SoHO \citep[e.g.,][]{1999Sci...283..810H, 2000JGR...10510419M, 2011ApJ...727....7M}, including the ubiquitous Alfvenic and magneto-convective energy input \citep[e.g.,][]{2007PASJ...59S.655D, 2007Sci...318.1574D, 2007Sci...317.1192T, 2012SSRv..172...69M}, and 2) in the transitions from the polar crown to the polar coronal hole environment we would sample the interface region between open and closed magnetic field to look for the signatures of interchange magnetic reconnection observed by Parker Solar Probe \citep[][]{1999ApJ...521..868F, 2020ApJ...894L...4F}.

\begin{figure}[!ht]
    \centering
    \includegraphics[width=0.8\textwidth]{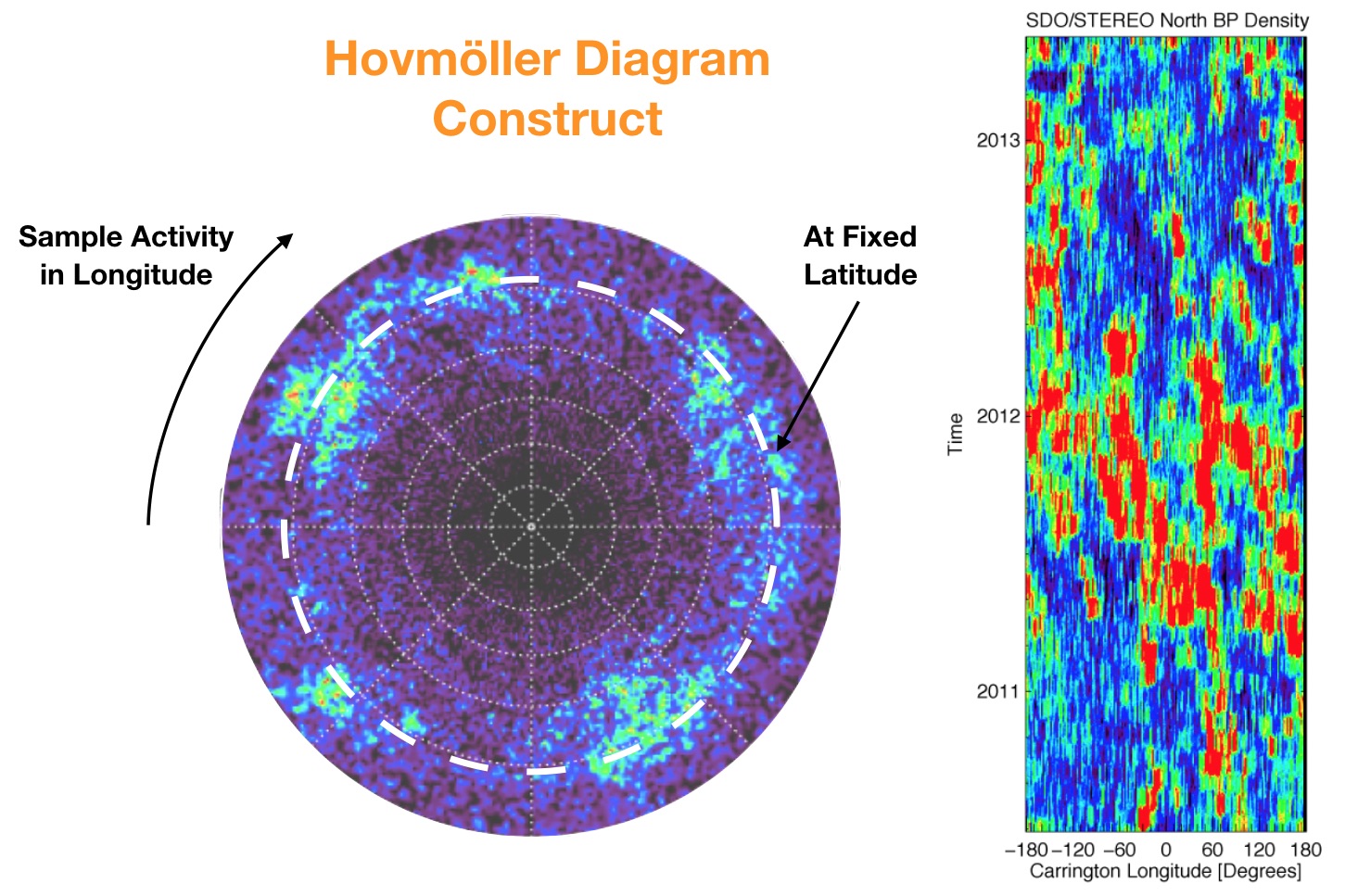}
    \caption{Illustrating the concept of a Hovm\"oller diagram. Left: For a synchronous spherical dataset a Hovm\"oller diagram is constructed by sampling the longitudinal evolution of the system in a narrow range of latitudes over time. Right: The annuli are then stacked to represent the passage of time. In the example shown the density of synchronous EUV BrightPoint density is shown and was the original example used to diagnose the presence of Rossby waves in the global-scale magnetic flux systems of solar interior.}
\label{fSMC5}
\end{figure}

The epoch from 2010 to 2014 created a unique opportunity in the space era \-- for the first time we had complete coverage of the solar disk in (extreme ultraviolet \-- EUV) observations provided by SOHO, SDO and the twin STEREO spacecraft. This unique set of observations permitted the study of the Sun's longitudinal behavior and to take a meteorological approach to the analysis of our star's atmosphere (see Fig.~\ref{fSMC5}). It was identified that certain longitudes remained active over many rotations, with those regions persistently producing a host of strongly eruptive behavior \citep[e.g.,][]{2015NatCo...6.6491M}. Further, in studying the behavior of small dipolar regions (EUV brightpoints) in the composite observations, it was discovered that the Sun's magnetic activity bands displayed the characteristic signature of (magnetized) Rossby waves \citep[][]{2017NatAs...1E..86M} that drove the longitudinal behavior of magnetic flux emergence and hence the key ingredients of space weather. Subsequent effort invested in understanding the nature of these waves \citep[][]{2020SpWea..1802109D} has demonstrated the potential of breakthrough by developing greater predictability of space weather phenomena and the magnetic activity in which it is rooted. A high latitude observing platform like HISM will have the opportunity to observe the development of these waves, their longitudinal structure and evolution and finally the organization of phenomena that they drive.

\subsection{Mission Science Objectives}
Using the examples presented above in concert with previous investigations of high-latitude and polar observing platforms \citep[]{POLARIS, 2005ESASP.592..663A, 2008nssv.book....1L, 2013AGUFMSH51D..02L, 2017arXiv170708193A, 2019AGUFMSH43F3352B} we formulate the following high level science goals for HISM that meet the those of the National Academy Decadal Survey ``Solar and Space Physics: A Science for a Technological Society'' \citep{NAP13060}.

\begin{enumerate}
\item Understand the Sun's internal structure and surface dynamics in the polar regions
\item Understand the 3D structure of the Solar/heliospheric magnetic field and its variation over time
\item Understand the variations in the solar wind speed and composition at high latitudes
\item Understand the origin and acceleration mechanism of solar energetic particles
\item Evaluate the use of high-latitude data for space weather predictions and warnings
\end{enumerate}

\label{sec:objectives}

\section{Science Instruments}

The instrument suite loosely based on the POLARIS study \citep{2008nssv.book....1L},
but descoped and updated where appropriate to meet the science 
objectives (Section~\ref{sec:objectives} and
achieve a feasible mission.
Their specifications and traceability to the science goals
are shown in Table~\ref{tbl:inst}. 


\begin{table}[bt]
 \caption{Instrument specifications, traceability, and requirements imposed on the
 spacecraft. The "Science Goals" column indicates flowdown from the
 science objectives listed in Section~\ref{sec:objectives}.}
\includegraphics[trim=0.7in 7.3in 0.7in 0.7in, clip, width=\textwidth]{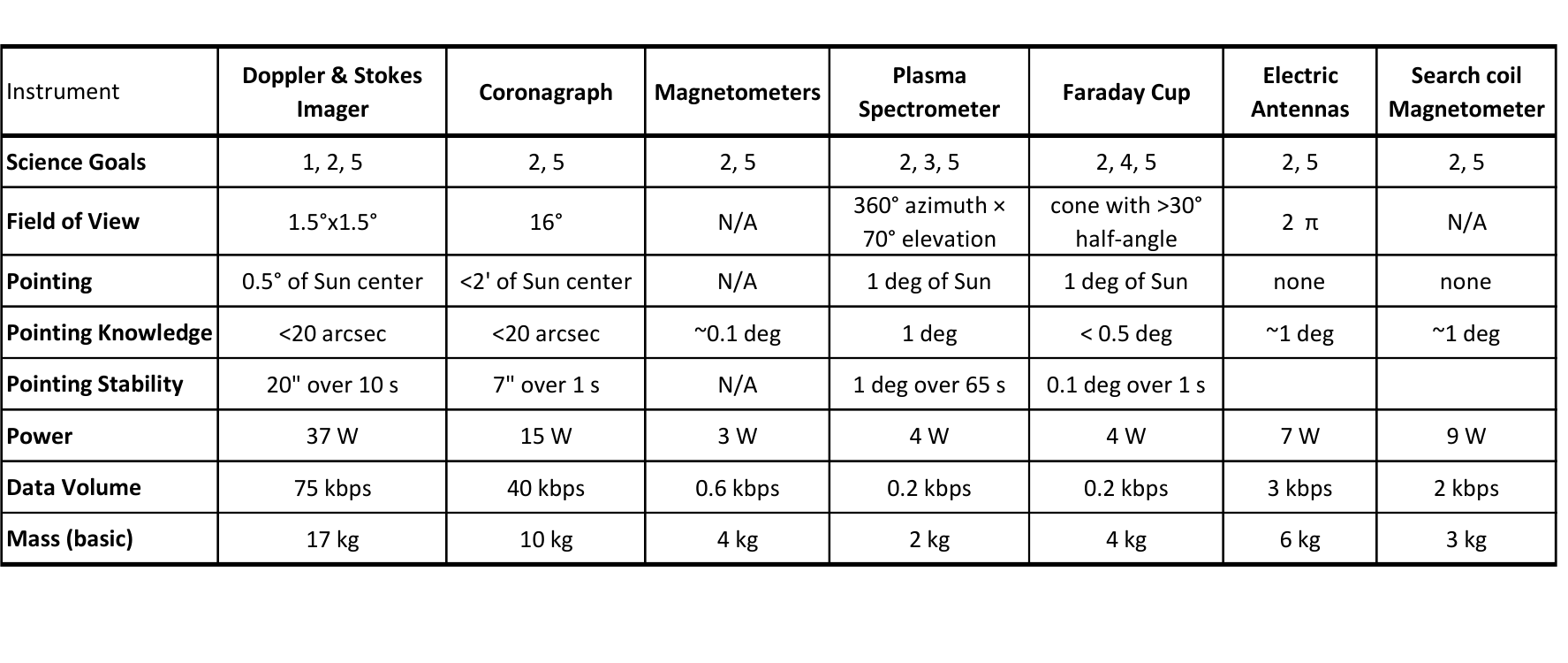}
 \label{tbl:inst}
\end{table}

\subsection{Doppler and Stokes Imager (DSI)}
The Doppler and Stokes Imager is used to measure the photospheric vector magnetic field structure near the Sun’s poles
and to probe the interior of the Sun, including its high latitude regions, through helioseismology.
The instrument is an imaging spectropolarimeter consisting of a telescope, polarizer and a narrowband tunable filter.
The instrument is based on the Heliospheric and Magnetic Imager (HMI) on the Solar Dynamics Observatory (SDO) satellite
as well as the Polarimetric and Helioseismic Imager for Solar Orbiter (SO/PHI). The field of view will be
adjusted to 1.5 x 1.5 degree, to allow full disk (~1 deg diameter) observation at 0.48 AU distance from the Sun.
The DSI mass is estimated as 15 kg total, based on preliminary instrument
designs by partner organizations (Jeffery Newmark, private communications).

 The instrument contains image stabilization system (tip/tilt mirror or similar)
 with a $>$20 arcsec stroke,
 but no roll correction. This imposes an instrument pointing
 requirement of 20 arcsec tip/tilt stability during exposure
 (up to 10s), and roll stability of also 20 arcsec over 10s, to insure $<$0.2” movement of the solar limb during exposure. 
 
The data rate is based on the assumption of 45 second cadence for velocity images and 5 minute cadence for Stokes parameter images, and lossless compression factor of 2. Helioseismology observations require long periods (days to weeks) of contiguous observations, so the SPI is required to be continuously sun-pointing except for flip maneuvers when crossing the equator; any design that requires periodic re-pointing of the spacecraft for data downlink is not allowed. 

\subsection{White Light Coronagraph}
The White Light Coronagraph is an externally occulted Lyot coronagraph, similar to the STEREO (Solar Terrestrial Relations Observatory) Sun Earth Connection Coronal and Heliospheric Investigation (SECCHI) COR2 and SOHO (Solar and Heliospheric Observatory)  Large Angle and Spectrometric Coronagraph (LASCO) C2 instruments. The primary purposed is to observe CME brightness and linear polarization. 

A coronagraph by its nature cannot have any optics in front of the occulter; the occulter and coronagraph
aperture must be aligned to the Sun to minimize scattered light. Because the Doppler and Stokes Imager has
a tip/tilt mirror, the coronagraph defines the overall pointing requirement for the Solar Polar Imager’s bus,
or its remote sensing instrument platform (instrument bus). The actual requirement is dependent on the
occulter diameter (inner edge of the field of view), which has not yet been determined from the science
goals. Following discussion with Joan Burkepile, for this study, the pointing accuracy requirement was
defined as 2 arcminutes, similar to the STEREO SECCHI COR2. The SECCHI COR2 has an inner
FOV of 2.5 R$_\cdot$, i.e. 1.5 R$_\cdot$ = 23’ between the solar limb and the inner FOV edge.
At 0.48 AU, the same 23’ margin translates to 0.75 R$_\cdot$, resulting in an occulter diameter of 1.75 R$_\cdot$. 
The size and mass are consistent with Appourchaux et al. 

\subsection{Magnetometers}
The magnetometer system is based on the Appourchaux studiy, with all specifications taken from that paper. The system consists of two laser-pumped vector helium magnetometers, mounted on a 5m boom, 1.5m apart. 

The magnetometer boom design is assumed to be the ATK (now Northrop Grumman) CoilABLE mast, used as magnetometer booms on most interplanetary and earth-orbiting satellites. Ideally the magnetometer boom should carry nothing but the magnetometers; however, DSCOVR mounted the solar wind instruments on the same boom as the magnetometers. Based on this precedent, and due to the crowded space around the SPI bus (plasma antennas, communication antennas, sail, radiators and solar arrays), the solar wind instruments were allowed to be mounted to the same boom.  

\subsection{Faraday Cup}
The Faraday Cup instrument is envisioned as a close copy of the
Solar Probe Cup (SPC), part of the Solar Wind Electron Alphas Protons
(SWEAP) suite on the Parker Solar Probe (PSP) spacecraft (Case et al., 2020).
The PSP/SPC is sun-viewing during the encounter phase between 0.25 AU down
to the eventual perihelion on 9.86 solar radii from the center of the Sun.
As such, the PSP/SPC was designed to withstand a very intense solar radiation
environment. For the HISM mission at 0.48 AU, some different material
choices and reduced radiation shields could allow for a reduction
in mass relative to the PSP application. A larger diameter of the
limiting aperture could allow for the Faraday Cup to have a
field of regard (FOR) of 45 degrees half-diameter and maintain
a large dynamic range for flux. The energy range for solar wind
ions would be about 100 eV to 6000 eV (corresponding to protons
with speeds from 139 to 1072 km/s). Electrons with an
energy/charge between 100 eV and 1500 eV can also be measured.
The variable command, AC operation, and synchronous detection of
the HISM/Faraday Cup would be the same as the PSP/SPC which allows
1-dimensional Velocity Distribution Functions obtained from one every
4 seconds to one every 250 ms. These time scales can be tailored to needs of the mission. 

\subsection{Plasma Spectrometer}
The Plasma Spectrometer is assumed to be identical to the
Fast Imaging Plasma Spectrometer (FIPS) on the MESSENGER spacecraft (Andrews et al., 2007),
which traces its heritage to the Solar Wind Ion Composition Spectrometer (SWICS)
instrument on the Advanced Composition Explorer (ACE). It consists of an
electrostatic analyzer and time-of-flight telescope, and is sensitive to
0.2 – 10 keV/Q solar wind ions. It has a field of view (FOV) of 1.4 pi
steradians, i.e. ~60 degree half-diameter cone. 
A nominal energy scan occurs every 65 seconds with a burst mode of 2 seconds
over a narrow energy range.  Mass resolution is such that C, N, and O are well separated.    

\subsection{Radio and Plasma Wave Package}
This package consists of two sets of sensors - a set of electrical monopole
antennas and a search coil magnetometer – and readout / data processing electronics.
The electrical antennas are assumed to consist of 3 antennas, each 6m long and 1.4 kg in mass,
based on the Solar Probe FIELDS antennas. These are deployed as far away from each other
as possible, while maximizing separation from other deployable structures and the sail.
The search coil magnetometer is assumed to mount on the magnetometer boom.
The electronics package is 5kg.

\subsection{Mission Requirements}

The driving requirements \& assumptions of the design are:
\begin{itemize}
    \item Accommodate science instruments as shown in Table~\ref{tbl:inst}
    \item Dedicated launch to $C_3=0$ trajectory
    \item Spiral-in to $0.48\,\mathrm{AU}$ circular ecliptic orbit, followed by cranking to
    $75^{\circ}$ inclination
    \item Utilize Solar Cruiser sail technology scaled up to $7000\,\mathrm{m}^{2}$
    \item Total mission life of $11$ years
    \item Maximize science observing time
\end{itemize}

The HISM mission has been designed by the NASA MSFC Advanced Concepts Office (ACO)
based on these requirements. Existing high-TRL components are used wherever possible.

\section{Spacecraft Design}

\subsection{Overview}

\begin{figure}[tbp]
\centering
\includegraphics[trim=1.3in 1.5in 0.5in 1.2in, clip, width=\textwidth]{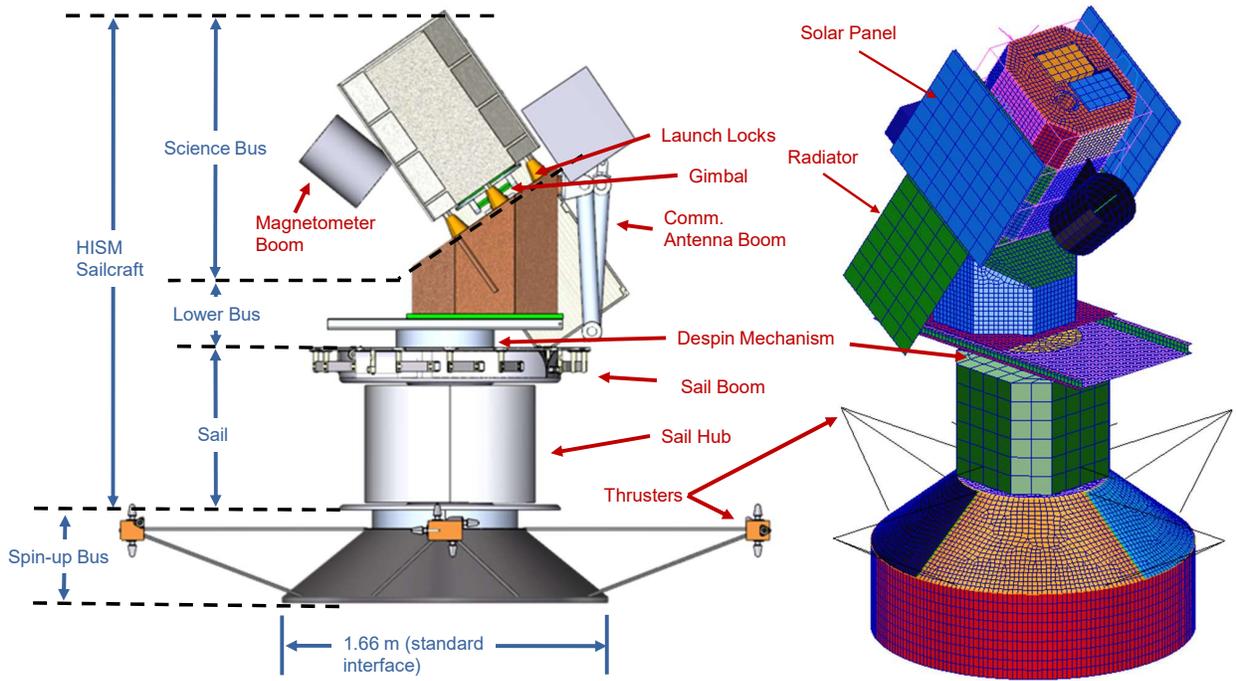}
\caption{Spacecraft in stowed configuration.}
\label{fig:stowed}
\end{figure}

\begin{figure}[tbp]
    \centering
    \includegraphics[trim=0.2in 1.2in 1.5in 1.2in, clip, width=\textwidth]{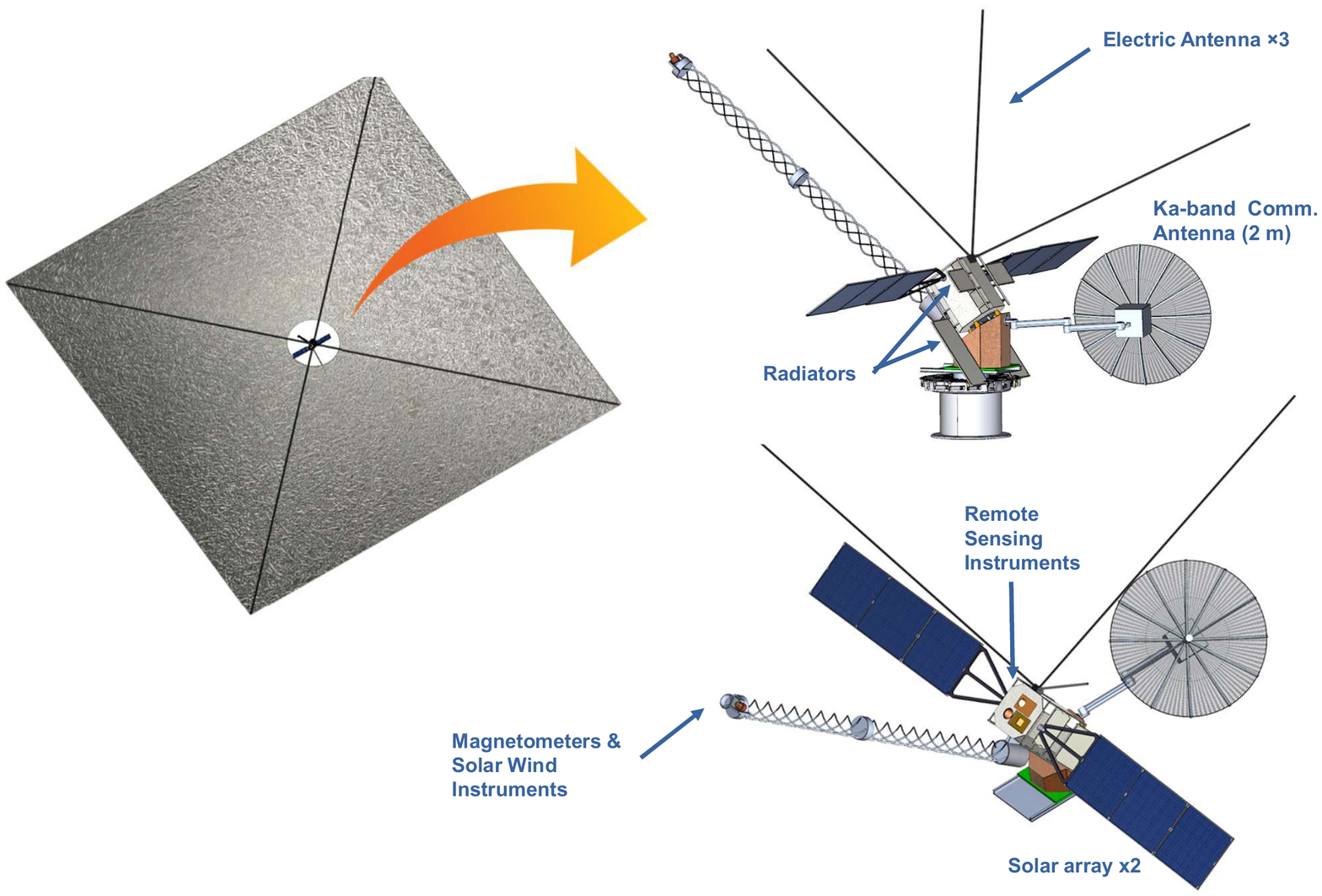}
    \caption{Sailcraft in fully deployed configuration, full sailcraft (left) and bus 
    detail (right). During the cranking phase of the mission, the sail is held at a constant
    35.3 degree angle relative to the Sun-spacecraft line. The sailcraft bus is designed
    with this angular offset, nominally pointing 35.3 degrees from the sail normal. The boom and
    electrical antennas are angled as far from the sail and from each other as possible.
    The sail has a $10\,\mathrm{m}$ hole in the center, which minimizes its effect on the
    science instruments and radiative coupling with the sailcraft bus.}
    \label{fig:sailcraft}
\end{figure}

The HISM spacecraft consists of the Science Bus, Lower Bus, sail assembly, and the
Spin-up Bus (Figures~\ref{fig:stowed},\ref{fig:sailcraft}).
The sail spins at $\sim 1$ rpm rate to allow centrifugal force to augment the stiffness 
(Section~\ref{sec:sail}).
Attached to the sail is the Spin-up Bus, which contains a propulsion system for
attitude control prior to sail deployment and for initiating the sail spin during sail deployment.
The Spin-up Bus is jettisoned after sail deployment;
the rest of the spacecraft is hereafter referred to as the sailcraft.
The Science Bus is a fine-pointing platform containing all science
instruments and their electronics, as well as the solar panels.
The Lower Bus contains the remaining avionics
subsystems, and while it is de-spun (decoupled from the rotating sail through a 
de-rotation mechanism), it is not required to meet the fine pointing requirements
of the science instruments. 

The mass of the sailcraft is estimated at 240 kg, or 293 kg including 
margin (mass growth allowance consistent with AIAA S-120A-2015). This results in 
sailcraft characteristic acceleration of  $Ac=0.265\,\mmss$ or $Ac=0.217\,\mmss$ respectively.
The total launch mass is 358 kg including margin.

\subsection{Sailcraft Bus}

The spinning solar sail is a major design constraint, as a spinning
platform is unsuitable for remote sensing observations.
One possible solution is to start science observations only after
the sailcraft has reached the  target high-inclination orbit and
the sail has been jettisoned.
Alternatively, a de-spin mechanism in the sailcraft can allow full
science operations while the sail is still attached and spinning.
The latter approach was chosen for HISM to 
maximize science returns by starting observations early in the mission
and continuing these observations from different orbital inclinations. 
This also allows the sailcraft to rely on Reflectivity Control Devices
(RCDs) for momentum management, eliminating the 
need for any propulsion system on the sailcraft for desaturating the momentum wheels.

Further, a large deployable structure such as a solar sail
cannot be assumed to be perfectly symmetric, and asymmetries
may lead to precession of the sail. 
Therefore the HISM bus design consists of 2 sections: the fine-pointing Science Bus
and the Lower Bus,
as shown in Figures~\ref{fig:stowed} and~\ref{fig:sail}.
The Lower Bus is de-spun from the spinning sail through a motorized
rotary joint, but is allowed to precess with the sail. The Lower Bus contains the
sailcraft avionics (section~\ref{sec:avionics}),
solar panel arrays, and the communication antenna boom.
The Science Bus contains all the science instruments, including the magnetometer boom,
and the science instrument electronics.
Each bus section has its own passive cooling system (Section~\ref{sec:thermal}).
The connection between the Science Bus and Lower Bus is through
a freely moving gimbal mechanism, with 3 degrees of freedom. Momentum wheels inside the
Science Bus keep the science instruments pointed at the Sun to the required pointing accuracy and
stability; when the momentum wheels are saturated, the gimbal mechanism is driven to the limit of the
motion range to transfer momentum to the sail, to be de-saturated using the RCDs.
The Science Bus is mounted at a nominal angle of $35.3^{\circ}$ angle relative
to the solar sail surface, corresponding to the optimal cone angle during the cranking phase of the mission.

\subsection{Spin-up Bus}
The Spin-up Bus is a modified launch vehicle adapter and fulfills that function
during launch. The bus also contains a propulsion system for initial attitude control 
and to spin up the sail during sail deployment. 
After the sail is deployed, the Spin-up Bus is jettisoned. To perform a safe jettison
and disposal, the Spin-up Bus contains a simple controller and a bare minimum of
navigation sensors (IMU and coarse sun sensor) which control the propulsion system to
fly away from the sailcraft.
The propulsion system is a blowdown hydrazine monopropellant system 
with 16 thrusters, each with 4N thrust and arranged into
4 pods of 4 thrusters each and located 1.75 m from the spacecraft centerline.

The total wet mass of the propulsion system is estimated to be 13.1 kg.
However, because the Spin-Up Bus is not part of the sailcraft,
the system may be replaced with a lower
efficiency (higher mass) system, such as cold-gas thrusters, without affecting the
sailcraft performance.

\subsection{Sail System}
\label{sec:sail}
The HISM concept is based on the Solar Cruiser mission, currently in Phase A development
at MSFC. The sail for the Solar Cruiser mission is
$1{,}666\,\mathrm{m}^2$ and provides a characteristic acceleration
($A_c$) of $0.17\,\mathrm{mm}/s^2$.
The sail architecture comprises a four
quadrant sail design with four composite
booms that are deployed from a central
rotating deployment mechanism and sail
hub. The sail membrane is the space- and
sail-proven aluminized CP1 polyimide
substrate successfully flown on NanoSail-D2
and to be flown on NEA Scout\citep{sobey_lockett}.

\begin{figure}[htbp]
    \centering
    \includegraphics[width=\textwidth]{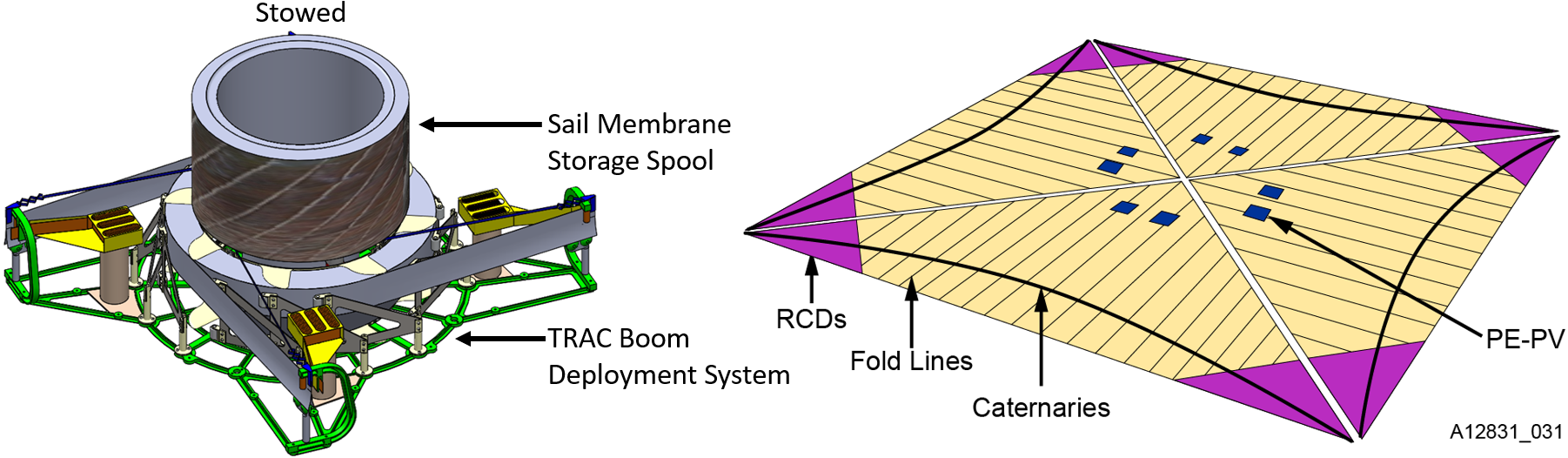}
    \caption{Solar Cruiser Solar Sail Stowed (left) and Deployed (right)}
    \label{fig:sail}
\end{figure}

The Solar Cruiser solar sail is pictured in both
the stowed and deployed configurations in
Figure~\ref{fig:sail}. The solar sail membrane is
deployed and tensioned using four
Triangular, Rollable, and Collapsible (TRAC)
high strain composite (HSC) booms, the
same geometry of the boom flight-validated
by NanoSail-D2 and used on the upcoming
NEA Scout mission.
TRAC booms have a triangular cross section
that flattens and rolls around a central spool
for stowage. Deployment
is actuated via a single motor controlled by
the bus. This proven design has higher
strength/weight for a given flattened height
than other rollable boom designs\citep{banik_murphey}
is easy to fabricate and taper, and has
flight heritage. The double omega boom was
rejected due to the manufacturing
difficulties for long booms, expensive
tooling, space constraints of the stowed system, and the sail application does not
require the additional torsional stiffness.

The HISM sail is equipped with Reflectivity Control Devices (RCDs, see section~\ref{sec:gnc})
for attitude control, along with Polyimide Embedded Photovoltaics (PE-PV) generating power
for the RCDs.
Both technologies will be demonstrated on the Solar Cruiser, further advancing the solar
sail architecture. 

Roccor conducted a feasibility study with a tentative goal of scaling the Solar Cruiser sail design 
to $A_c=0.3\,\mathrm{mm}/\mathrm{s}^2$
with a payload (non-sail-subsystem) mass of $140\,\mathrm{kg}$. 
The sail membrane was assumed to be the same design as Solar Cruiser
($2.5\,\mu\mathrm{m}$ CP1 substrate), and the
sail system configuration was assumed to be the same as well,
i.e.\ 4 sail quadrants supported by
4 tapered composite TRAC booms on a common spool for deployment. 
For a rigid or non-spinning sail system such as Solar Cruiser, each boom is under
compressive stress from sail tension, and therefore the driving constraint for sizing of
the system is the buckling of the booms. The tension requirement itself is derived
from the stiffness requirement of the deployed system,
i.e.\ maintaining a sufficiently high first mode natural frequency,
as well as the flatness requirement of the sail membranes.
As sail sizes become very large, the size and mass of the booms required
to tension the sail become disproportionately large.
The study concluded that, in order to meet the HISM requirements, sail tension
would need to be imparted using centrifugal force by spinning the sail
at $\sim$1rpm. This centrifugal force reduces the buckling strength
requirement of the composite TRAC booms, significantly decreasing their mass.
Since the booms are no longer required for sail tension, their primary function
then becomes to provide out of plane stiffness to the sail membrane to improve
sail-flatness and to reduce attitude control complexities experienced by other
spinning sail missions such as
IKAROS\citep{yuki}.
The resulting design achieves a $52\,\mathrm{kg}$ sail subsystem mass with a $7000\,\mathrm{m}^2$ sail area, including 4 booms, sail membranes, 
stowage spools, deployment mechanism, and $3\,\mathrm{kg}$ allowance for RCDs.

\subsection{Structures}

The HISM structure has been designed to meet strength and stability requirements in
NASA-STD 5001B, assuming a Falcon-9 launch vehicle. A conventional all-metal construction
using 2000 series aluminum and Hexcel aluminum core is assumed. 
The structural model is shown in figure~\ref{fig:stowed} (right). The structural mass of the
science bus and  lower bus are estimated to be
22.2 kg and 23.2 kg respectively.


\subsection{Command and Data Handling}

\label{sec:avionics}

The spacecraft avionics performs all command and data handling
for the spacecraft, and perform the data storage and downlink operations for the
science instruments.
It is a 10-board stack of PCI104s form factor avionics boards, consisting of:
a Single Board Computer (SBC) for flight control, 
a digital signal processor board for data management and formatting of instrument data,
a digital I/O board and an analog I/O board, three memory storage boards,
a reaction wheel controller/driver board, an antenna gimbal controller/driver board,
and the avionics stack power supply board.
Mass and power estimates are based on currently available components with flight heritage,
such as the Space Micro Proton 400k SBC. 
The memory boards provide 8GBytes each, for a total of
24GBytes of on-board memory storage. The predicted mass of the avionics system is 9.7 kg, including cabling, and power consumption is 84.3 W.

\subsection{Communications}

The HISM communication system consists of an X-band system with non-directional
patch antennas and a Ka-band system using a 2-meter antenna. 
The X-band system is used for initial checkout and sail deployment.
The Ka-band system is used for all subsequent communications, while the X band
system remains available as a backup.

The X-band system
consists of a Tether Unlimited SWIFT XTX 7-watt transceiver,
4 patch antennas on the sailcraft bus, and an additional patch antenna on the Spin-up Bus,
allowing uninterrupted
communications in all directions during sail deployment.
During sail deployment, expected to occur at a distance of
400k to 600k km from Earth, the X-band system is capable of
2 Mbps downlink to a DSN 34m antenna with 54 G/T. This allows for 
1 fpm uncompressed full-HD video from two onboard cameras,
or a higher rate if compression is used.

During the science observing phase
(cranking phase) of the mission, science data is generated at an
average rate of 275 kbps (Table~\ref{tbl:inst}).
With 4:1 compression for science data, 2.5 kbps spacecraft telemetry and 
30\% margin, 
the data volume is 8.0 Gbits/day. 
A Ka-band system with a 2-meter antenna 38W RF power has been selected as an optimal 
balance between DSN cost, RF power and antenna size.  This configuration is
capable of 1 Mbps downlink at the mission's maximum Earth-spacecraft distance of 220 Gm,
requiring 2.2 hours of downlink time per day at this distance. 
A General Dynamics (GD) Small Deep-Space Transponder (SDST) and
along with 
Bosch Ka-band AstralK Traveling Wave Tube Amplifier (TWTA) are assumed for this study.
%

Because the required sail orientation and Sun-pointing science instruments constrain
the spacecraft attitude, the Earth can be almost any direction relative to the spacecraft. 
A 5-meter long articulated, gimbaled antenna boom is used to insure
communications throughout the mission.
A 2-axis gimbal unit is placed at the mount of the antenna to the boom,
while the boom its self has 3 axes of movement: two at the boom to S/C union, and one at
the boom joint.
The antenna boom system is estimated at 11 kg, including 2.5 kg for actuators and
6 kg for the 2-meter mesh antenna itself.

\subsection{Guidance, Navigation and Control}
\label{sec:gnc}

The GNC subsystem has some challenges for the HISM mission. The large sail and
high spin rate lead to a large amount of angular momentum stored in the sail. This removes
traditional attitude control actuators (RCS thrusters or reaction wheels) from the trade space for
controlling the sail attitude. To do coarse steering of the vehicle, 
RCDs and active mass transfer (AMT) were both studied for HISM. 

RCDs are liquid-crystal
devices that switch between transparent and translucent (diffuse) states based on the 
applied voltage. This principle has beeen successfully
tested on the IKAROS mission \citep{RCD} and is also currently under development by
Nexolve for the Solar Cruiser. A set of 8 RCD panels, each with a
$4.6\,\mathrm{m}^2$ area and located
at the outer corners of each sail quadrant, is baselined for the HISM design.
This allows for $10^{\circ}/\mathrm{day}$ slew rate in the 0.48 AU orbit.
Following the Solar Cruiser design, the RCDs are placed at an
angle relative to the sail plane, allowing for roll torque to be generated by selectively turning 
on/off the RCDs.
Each RCD is powered by sail-embedded PE-PV panels, currently under development
at MSFC \citep{LISA}. The PV cells are located near the RCDs, along with a controller, to minimize
cable mass and transmission loss. The controllers are commanded from a wireless transmitter in the sailcraft bus.

The active mass transfer (AMT) device is a 2-axis translation mechanism
that shifts the center of mass (CM) of the sailcraft bus relative to the center of pressure (CP)
of the sail, generating a pitch or yaw torque. The baseline HISM design uses only RCDs for control of the sail; however, an AMT can be added if a decrease in RCD area is necessary or for redundancy in
the system. A slew rate of $10^{\circ}/\mathrm{day}$ is well within the capability of an AMT with 
$\pm 21.5\,\mathrm{cm}$ motion range in each axis.

When the sail is deployed, the lower bus will be de-spun through a motor at the sail/bus
interface. For attitude knowledge of the sail, the lower bus contains a coarse sun sensor and a
star tracker.
The science bus contains a star tracker and a more precise sun sensor for fine
pointing and pointing knowledge of the science instruments.
Three 0.10 NM reaction wheels are contained in the Science Bus to achieve the science instrument
pointing requirements (Table~\ref{tbl:inst}), using input from the fine sun sensor.
An IMU is also be included for more accurate rate measurements.

\subsection{Power}

The HISM sailcraft is designed for operation from 1 AU to 0.48 AU, with the solar irradiance varying by
a factor of 4.3. And while full science operation is not planned till the 0.48 AU orbit, the sailcraft
requires significant power for survival heaters at 1 AU, owing to the large radiators necessary for 
operation at 0.48 AU.
As a consequence, the HISM power system is designed to provide a maximum of 700 W at 1 AU
and 540 W at 0.48 AU. 

The primary solar arrays on HISM are two "reflective" solar panels, with a total area of $1.7\,\mathrm{m}^2$ 
and coated with a pattern of highly reflective gold coating, covering
60\% of the total area. This coating reduces solar input to the panel, allowing them to operate at 0.48 AU and
generate 540W. These are supplemented by two conventional solar arrays, with a total area of $2\,\mathrm{m}^2$,
generating 458W at 1 AU. At 0.6 AU distance from the Sun, the conventional arrays will be electrically
disconnected from the main arrays. 
The solar arrays are grouped into two deployable arrays. The total mass of the power system, including the panels,
control board, power switch board, and harness, is estimated at 20.7 kg.

The sailcraft does not contain any batteries. The power for initial checkup, attitude control,
solar array deployment and thruster operation (including propulsion preparation heating)
is supplied by
a 205 Whr Li-Ion secondary battery inside the Spin-up Bus, to be discarded after sail deployment.


\subsection{Thermal Design}
\label{sec:thermal}

The HISM sail is CP1 film with aluminum coating on the Sun side and black chromium coating on the
back side. Black chromium has an emissivity of $\sim 0.4$; the sail membrane temperature is estimated
to be $120^{\circ}\mathrm{C}$ at 0.48 AU from the Sun,
well below the CP1's temperature limit of $250^{\circ}\mathrm{C}$. A $10\,\mathrm{m}$ hole in the
center of the sail minimizes radiative coupling between the sail and the sailcraft bus.
Black chromium is also electrically conductive, which minimizes electrostatic charging of the sail.

The HISM sailcraft is designed for full science operation at 0.48 AU distance from the Sun.
The science bus and lower bus are both enclosed in high-temperature multi-layer insulation (MLI),
composed of multiple layers of embossed Double Aluminized Kapton with an outer layer of single
aluminized Kapton. Other spacecraft elements, such as radiator panels, high-gain antenna
and Solar sail booms and deployment housing are covered with low-absorptivity white paint.
Many smaller spacecraft elements are covered with low-absorptivity white paint,
high-temperature MLI or other thermal coatings as necessary.  

The Science Bus and Lower Bus are each equipped with a pair of deployable heatpipe radiator panels oriented
edge-on to the Sun.
Heat from internal components are transported to the radiator panels using a network of
heatpipes and heatpipe-embedded cold plates. 
The mass of these control systems is estimated to be 17.8 kg for the Science Bus and 19.9 kg for the Lower Bus.

Because the passive cooling system is designed for full science operation at 0.48 AU,
the bus temperatures will fall below the assumed $-20^{\circ}\mathrm{C}$ survival temperatures early 
in the mission. 
Survival heaters are used to keep the internal components above the survival temperatures;
a total of $310\,\mathrm{W}$ power heater power is required at
1 AU with science instruments turned off.

\section{Mission Design}

\subsection{Mission Profile}

HISM is launched on a dedicated launcher. With a launch mass of 358 kg and
the requirement to be launched into a $C_3>0$ trajectory, many existing launch
vehicles are suitable for this mission, such as the Falcon-9 or Atlas-V 401.
Shortly after the spacecraft (sailcraft + Spin-up Bus) is separated from the
launcher, the thrusters in the Spin-up Bus are used to de-tumble the spacecraft
and establish attitude control, and the solar arrays on the Lower Bus
are deployed. The sail is then deployed, using the Spin-up Bus thrusters
to increase the spin rate of the sail and ensure a sufficient spin rate to
maintain the sail's structural integrity. After the sail is fully deployed,
the Spin-up Bus disconnects from the sailcraft and uses the thrusters to
fly away from the sail. The sailcraft then begins the inward spiral phase of the
mission (Section~\ref{sec:mission_analysis}). The sailcarft remains in the ecliptic plane
while it reduces the orbital radius from 
1 AU to 0.48 AU. 

Once a circular orbit is achieved at 0.48 AU, all science instruments will start continual
operations. The sailcraft switches from spiraling to cranking
operation, where the sail attitude is optimized for orbital inclination change.
The orbit is a constant distance from the Sun, and the sail cone angle will
also be held constant; only the sail clock angle (roll angle around the
sailcraft-Sun line) will change.
From this point on, the sailcraft continues to increase its orbital
inclination (Figure~\ref{fig:orbital_param}).
There is no defined end to the cranking phase; with no onboard consumables and assuming the
sailcraft remains operational, it will eventually reach $90^{\circ}$ inclination.
At this point the sailcraft may continue cranking past $90^{\circ}$, or 
allow its orbital inclination to oscillate around $90^{\circ}$.


\subsection{Mission Analysis}
\label{sec:mission_analysis}

\begin{figure}
    \centering
    \includegraphics[width=0.9\textwidth]{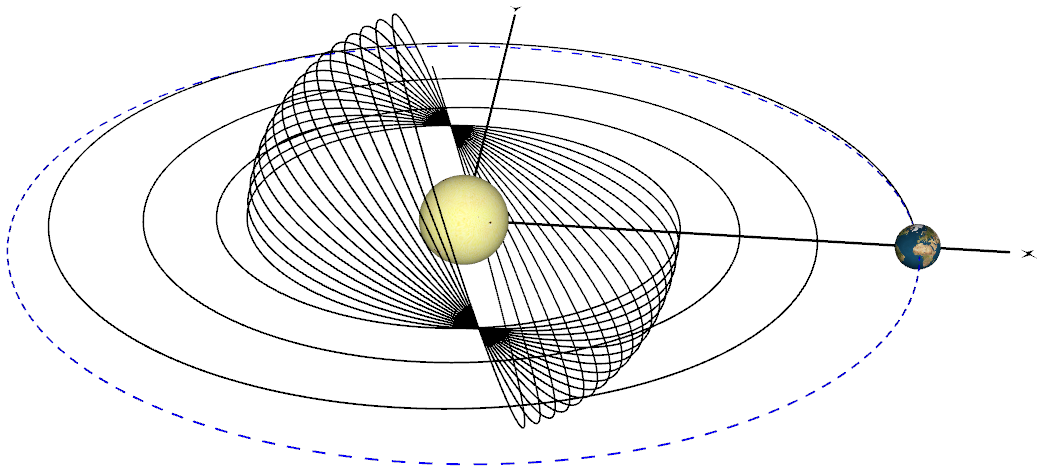}
    \caption{Solar Sail Mission Trajectory from Earth (C3 = 0) to 0.48 AU and 75 deg Inclination for an Example Sail Characteristic Acceleration of 0.30 $\mmss$}
    \label{fig:trajectory}
\end{figure}

Minimum time trajectories for the inward spiral phase and cranking phase
were generated for various sail characteristic
accelerations, $A_c$, which is a sailcraft performance metric that 
represent the thrust accelerations that sails can achieve
at a solar distance of 1 AU with the sail perpendicular to sunlight. For each case,
the sail departs Earth with a C3 = 0 and performs an inward spiral
to the 0.48 AU solar distance. The sail then begins the "cranking" phase of the mission, during which the sail increases the orbital inclination while maintaining a constant distance from the Sun. Figure~\ref{fig:trajectory} shows the inward spiral and cranking phases for a sail $A_c$ of 0.3 $\mmss$.

Trajectories for the spiral and cranking phases were each optimized separately. For the spiral-in phase, 
a minimum time trajectory was calculated with initial conditions of $C_3=0$ (i.e. launched from Earth at 
escape speed), $r=1\,\mathrm{AU}$ and $v_r=0$. The targeted final conditions were $r=0.48\,\mathrm{AU}$, $v_r = 0$ and $dv_r/dt=0$. The cone angle
(angle between the sail normal
and sunlight) was allowed to vary for optimization, while the clock angle was held at 
For the cranking phase, the clock angle was held at 270$^{\circ}$ to maximizes the lateral component of thrust. During the cranking phase, the cone angle is held constant at 35.3 $^{\circ}$, while the clock angle is varied to maintain the constant solar distance.
The sun incidence angle is required to change from $35.3^{\circ}$ towards ecliptic north to 
$35.3^{\circ}$ towards ecliptic south and vice versa for each orbit.  
These two maneuvers were assumed to be instantaneous for the purposes of this study. 

\begin{figure}
    \centering
    \includegraphics[trim=1in 3.3in 1in 1.7in, clip, width=0.8\textwidth]{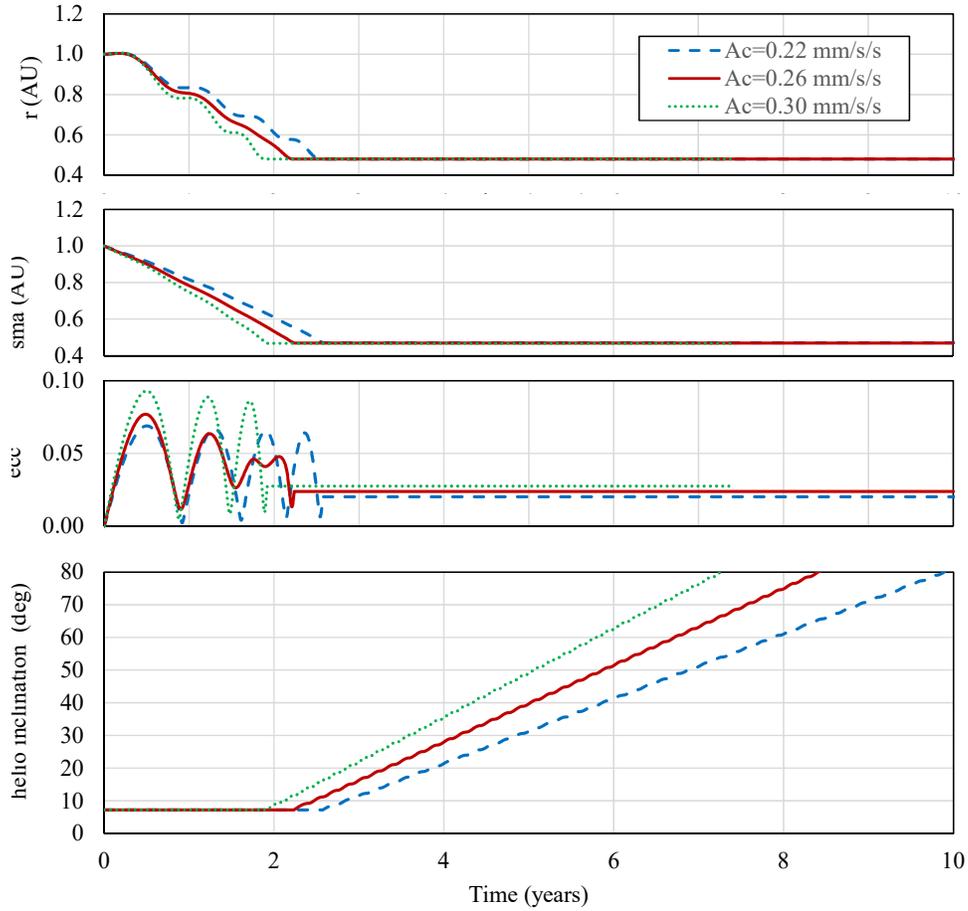}
    \caption{Spacecraft distance from Sun -- along with orbital semi-major axis, eccentricity and inclination (relative to the 
    Sun's equator) -- plotted for 3 different characteristic accelerations.
    $A_C=0.26\,\mathrm{mm}/\mathrm{s}^{2}$ corresponds to the HISM design and predicted mass (without margin); at
    this characteristic acceleration,
    duration from launch to 0.48 AU is 2.2 years. Heliographic inclination of
    $60^{\circ}$ is reached 6.7 years after launch, and $75^{\circ}$ at 8.0 years.
    }
    \label{fig:orbital_param}
\end{figure}



Minimum time trajectories were generated over a range of characteristic accelerations. 
Figure~\ref{fig:orbital_param} shows the orbital parameters
throughout the mission; $A_c=0.22\,\mathrm{mm}/\mathrm{s}^2$ and
$0.26\,\mathrm{mm}/\mathrm{s}^2$ represent
characteristic accelerations based on the mass estimates from this study -- with and without mass margin, respectively. 
The $A_c=0.3\,\mathrm{mm}/\mathrm{s}^2$ case is also displayed for comparison purposes.  Mission durations are shown in 
Table~\ref{tbl:transit_time}.

\begin{table}[htb]
    \centering
    \caption{Times to Key Mission Events for Sailcraft with Characteristic Accelerations of
    0.22, 0.26, and 0.30\,$\mmss$. Total times are given to 60$^{\circ}$ and 75$^{\circ}$ Heliographic Inclinations}
    \begin{tabular}{|c|c|c|c|}
         \hline
          &  $A_C = 0.22\,\mmss$ & $A_C = 0.26\,\mmss$ & $A_C = 0.30\,\mmss $ \\
         \hline
         Time to 0.48 AU [yr] & 2.6 & 2.2 & 1.9 \\
         Total time to $60^{\circ}$ [yr] & 7.9 & 6.7 & 5.8 \\
         Total time to $75^{\circ}$ [yr] & 9.4 & 8.0 & 6.9 \\
         \hline
    \end{tabular}
    \label{tbl:transit_time}
\end{table}

\section{Conclusions}

A spacecraft in high inclination ($>60^{\circ}$) orbit, equipped with a combination
of remote sensing and in-situ instrument, has the potential to dramatically
change our understanding of the Sun and the heliosphere. The HISM concept study
demonstrates the feasibiilty of such a 
mission using currently available components and technologies, and by using the solar sail 
technology currently under development for Solar Cruiser.
The mission accommodates
46 kg of science instrument payload (60 kg including margin), including remote sensing instruments
requiring 2 arcminute pointing accuracy and 7 arcsec/second stability.
The sailcraft basic mass, estimated based on currently available components, is 
estimated at 240 kg,
achieving a characteristic acceleration of  $0.26\,\mmss$ which
is sufficient to reach a $60^{\circ}$ inclination (heliographic) in 6.7 years from launch.
Because the sailcraft has no onboard consumables, the maximum inclination and mission life
are only limited by operations funding and longevity of the sailcraft components. 
With mass margin, and absent further improvements in available components, the
characteristic acceleration is $0.22\,\mmss$, requiring 7.9 years from launch to $60^{\circ}$.

\bibliographystyle{abbrvnat}
\bibliography{hism_bibliography}  


\end{document}